\documentclass{aa} % for a paper on 1 column
\usepackage{natbib}
\usepackage{graphicx}
\usepackage{txfonts}
\usepackage{epsfig}
\usepackage{times}
\usepackage{rotating}
\usepackage{float}
\usepackage[caption = false]{subfig}
\usepackage{setspace}
\usepackage{lscape}
\usepackage{epstopdf}
\usepackage{tabularx}
\usepackage{caption}
\usepackage{lineno}
\usepackage[usenames]{color}
\usepackage[toc,page]{appendix}
\usepackage{soul}%Per testo barrato, togliere alla fine
\usepackage{amsmath}

\begin{document}

%\title{Galaxies stripped before entering the cluster: Evidence of pre-processing in the Virgo outskirts}
\title{Hierarchical assembly  in action: a galaxy tail from
a disrupting group in the Virgo cluster outskirts.}
\author{J. Alfonso L. Aguerri\inst{1,2}, Stefano Zarattini\inst{3}, Virginia Cuomo\inst{4} \& Lorenzo Morelli\inst{5} }

\institute{Instituto de Astrof\'isica de Canarias, calle Vía L\'actea s/n, E-38205 La Laguna, Tenerife, Spain
\and Departamento de Astrof\'isica, Universidad de La Laguna, Avenida Astrof\'isico Francisco S\'anchez s/n, E-38206 La Laguna, Spain
\and Centro de Estudios de F\'isica del Cosmos de Arag\'on (CEFCA), Plaza San Juan 1, 44001 Teruel, Spain
\and Departamento de Astronom\'{\i}a, Universidad de La Serena.Avenida Ra\'ul Bitr\'an 1305, La Serena, Chile
\and Instituto de Astronom\'{\i}a y Ciencias Planetarias. Universidad de Atacama. Avenida Copayapu 485, Copiap\'o 1530000, Atacama, Chile.\\
jalfonso@iac.es}

\date{\today}

\abstract
{Group environments are thought to play a key role in shaping galaxy evolution prior to cluster accretion. However, direct observational evidence linking group--cluster interactions to the transformation of low-mass galaxies remains scarce.}
{The nature and origin of the W cloud, located in the southern outskirts of the Virgo cluster, have been reexamined in order to better understand the dynamical processes driving group accretion and galaxy transformation during cluster assembly.}
{A comprehensive analysis combining the spatial distribution, kinematics, and stellar population properties of galaxies in the W cloud and its surroundings reveals the dynamical structure of this system. This multidimensional approach allows us to characterize the three-dimensional structure of the system, assess its dynamical state, and identify the dominant environmental mechanisms at work.}
{We conclude that the W cloud is not a filament of the large-scale structure seen in projection. Instead, the W cloud is dominated by a compact galaxy group (the W group) currently interacting with the main cluster. Moreover, a newly discovered, dynamically coherent tail of galaxies (the W tail) connects the W group to the cluster and exhibits a continuous sequence in velocity, velocity dispersion, and three-dimensional distance. The low-velocity component of the tail is already gravitationally bound to Virgo, while higher-velocity galaxies remain dynamically connected to the group and are still infalling. The W tail forms a planar structure aligned with the orbital geometry of the W group, strongly indicating a tidal origin. The stellar mass and color properties of their members show that the stripped population is dominated by low-mass, star-forming dwarfs that remain in the blue cloud, with their slightly bluer colors relative to the W group naturally explained by their lower stellar masses rather than by environmentally induced star formation enhancements.}
{The W group--W tail system provides a well-resolved example of an active group--cluster interaction, illustrating how low-density galaxy groups can deliver largely unprocessed dwarf galaxies into clusters. This system offers key observational constraints on the hierarchical assembly of galaxy clusters and the buildup of their dwarf galaxy populations.}


\authorrunning{J. A. L. Aguerri et al}
\titlerunning{A galaxy tail from a disrupting
group in the Virgo cluster outskirts}
\maketitle
\nolinenumbers

%------------------------------------------
%-----------------------------------------
%-----------------SECTION 1---------------
%-----------------------------------------
%------------------------------------------
\section{Introduction}
\label{sec:intro}

On megaparsec scales, the distribution of galaxies in the Universe is far from homogeneous \citep{Sarkar2009}. This large-scale structure arises naturally from gravitational evolution: the anisotropic collapse of the primordial density field gives rise to the intricate network of clusters, groups, filaments, walls, and voids that make up the cosmic web \citep{Zeldovich1970, Bond1996}. These structures contain very different fractions of the total mass and volume of the Universe, and galaxies continuously migrate through them over cosmic time \citep{Cautun2014}. Each environment leaves distinct imprints on galaxy properties, making the influence of the environment a key ingredient in understanding galaxy formation and evolution \citep{Boselli2006}.

Traditionally, the influence of environment on galaxy evolution has been studied by comparing the properties of galaxies in high-density regions (clusters or groups) with those in low-density, field environments. These studies revealed a clear environmental dependence: galaxies in dense environments are dominated by red, early-type systems, while those in low-density regions are typically blue, late-type, and actively star-forming \citep{Dressler1980, Balogh2004, Aguerri2007, Peng2010}. These morphology–density and color–density relations provide strong evidence that environment plays a fundamental role in driving galaxy transformation.

A variety of physical mechanisms have been proposed to explain these transformations during the infall of galaxies into dense environments. These include tidal interactions and harassment \citep{Moore1996, Moore1998, Mastropietro2005, Aguerri2009, Boselli2014}, major and minor mergers \citep{Toomre1977, Aguerri2001}, ram-pressure stripping by the intracluster medium \citep{Gunn1972, Quilis2000, Boselli2006}, and the gradual removal or exhaustion of gas reservoirs through strangulation or starvation \citep{Larson1980}. These mechanisms act with varying efficiencies depending on the local density and the depth of the gravitational potential, collectively leading to the quenching of star formation and the morphological transformation of galaxies from disk-like to spheroidal systems.

The advent of large spectroscopic surveys such as the Sloan Digital Sky Survey \citep[SDSS;][]{York2000} and the Dark Energy Spectroscopic Instrument \citep[DESI;][]{DES2016} has opened a new era in environmental studies. The vast statistical samples provided by these surveys allow the characterization of galaxy properties across the full range of cosmic environments, from isolated voids to massive clusters, and have enabled the detection and mapping of filaments and other large-scale structures in unprecedented detail \citep[e.g.,][]{Tempel2014, Chen2016, Rost2020}. These data now make it possible to study how galaxies evolve as they migrate along the cosmic web, from voids into filaments, and eventually into clusters.

 Recent works have shown that the influence of environment begins well before galaxies reach the virialized regions of clusters. Evidence from both observations and simulations indicates that galaxies already experience significant transformations within filaments and groups, a process often referred to as pre-processing \citep{Fujita2004, Dressler2004, Haines2015}. This includes early quenching of star formation \citep{Rojas2004, Kuutma2017, Luber2019}, morphological changes \citep{Kuutma2017, Castignani2022a, Castignani2022b, Cuomo2026}, and gas depletion occurring prior to cluster infall \citep{Aragoncalvo2019,Zarattini2025}. This is mainly because on filaments there is a significant increase in the galaxy overdensity with respect to the field, but smaller than in clusters or groups \citep{Aguerri2026}. Consequently, the impact of environment on galaxy evolution appears to be continuous throughout a galaxy’s lifetime, rather than confined to the final stages of cluster accretion.

Within the current cosmological framework, galaxy clusters assemble hierarchically through the continuous accretion of galaxy groups and field systems along the filaments of the cosmic web \citep[e.g.,][]{Springel2006, DeLucia2007, Haines2018}. These infalling structures contribute not only to the growth of the cluster dark matter halo, but also to the progressive buildup and transformation of its galaxy population. The outskirts of clusters, therefore, represent key regions where galaxies experience the cumulative impact of multiple environmental mechanisms, bridging the gap between group-filament pre-processing and cluster-driven evolution \citep{Reiprich2013, Gill2005}

Several studies of nearby galaxy clusters have highlighted this crucial role of their outskirts in shaping galaxy evolution. Observations of clusters such as Coma, Fornax, and Abell 1367 show that galaxies in the infall regions and along filaments are subject to a combination of environmental effects, including tidal interactions, galaxy harassment, and preprocessing within groups, before reaching the cluster core \citep[e.g.,][]{Haines2015, Jaffe2016, Boselli2022}. These regions often host infalling groups that are partially disrupted, contributing not only to the growth of the cluster's dark matter halo but also to the assembly of its galaxy population \citep[e.g.,][]{Fujita2004, DeLucia2012}. Moreover, studies combining kinematics, stellar populations, and gas content reveal that galaxies in the outskirts can exhibit gradual quenching, morphological transformation, and gas depletion, indicating that cluster-driven processes begin to act well before galaxies reach the virialized region \citep[e.g.,][]{Wetzel2013}. These results underscore that the outskirts of clusters are laboratories for studying hierarchical accretion and environmental preprocessing, providing direct insight into the continuous assembly of clusters and their galaxy populations.

The Virgo cluster provides an exceptional case for investigating the hierarchical growth of galaxy clusters and the progressive assembly of their galaxy populations. As the nearest massive cluster to the Local Group (LG), Virgo offers the opportunity to study galaxy properties across a wide range of masses and environments, from giant ellipticals to the faintest dwarfs \citep{Binggeli1985, Lisker2007, Ferrarese2020}. Its infall region exhibits a rich filamentary structure, with numerous galaxy groups and filaments feeding the cluster \citep[e.g.,][]{Ftaclas1984, Binggeli1993, Kim2016, Boselli2018, Castignani2022a, Castignani2022b}. The outskirts of Virgo constitute a nonrelaxed environment where several already known galaxy aggregations, such as the M, L, and W clouds \citep{deVaucouleurs1973, Boselli2018}, are currently interacting with the cluster and represent active sites of galaxy accretion and transformation. Studying these systems provides direct insight into the ongoing assembly of Virgo and the environmental processes shaping its galaxy population.

This work is focused on the outskirts of the Virgo cluster and provides a unique observational view of the hierarchical cluster assembly in action.  Within the well-studied W cloud, in this work we identify three new structures. The first one is what we called the W group, a group of galaxies with coherent properties that are found in the heart of the classical W cloud. The second structure is a large area of the sky, partially overlapping the Virgo Southern Extension \citep[SE: ][]{Tully1982} in which we find that several galaxies were found at the same distance from the Virgo cluster center as the galaxies of the W group. We call this area the W plume. Within this region, we finally identify a subset of galaxies that not only fulfill the distance criterion but also have velocity, velocity dispersion, orbital geometry, and colors compatible with those of galaxies stripped from the recently introduced W group. We call this structure the W tail. The W group and W tail are thus new structures, linked with one another, and are properly defined in Sect. 2.5. We are witness to a snapshot of this ongoing buildup, allowing us to study the physical mechanisms responsible for galaxy transformation before and during cluster infall. By capturing this process in real time, our study offers a rare opportunity to connect the large-scale dynamics of group–cluster interactions with the evolution of individual galaxies, providing critical insights into how clusters grow and how their dwarf populations are assembled. In this paper we used the cosmological model with $H_{0} = 70$ km s$^{-1}$ Mpc$^{-1}$, $\Omega_{m} = 0.3 $ and $\Omega_{\Lambda} = 0.7$.

\section{Spectroscopic catalog of the Virgo cluster}
\label{sec:catalog}

In this section, we describe the spectroscopic catalog of the Virgo cluster employed in this study, including the velocity corrections and coordinate transformations applied to the data. We compute supergalactic coordinates and define the selection of galaxies belonging to the W cloud and the W plume, which provides the foundation for the subsequent analysis.

\subsection{The Virgo cluster}

\begin{figure}
\centering
\includegraphics[width=0.5\textwidth, trim = 40 20 70 30]{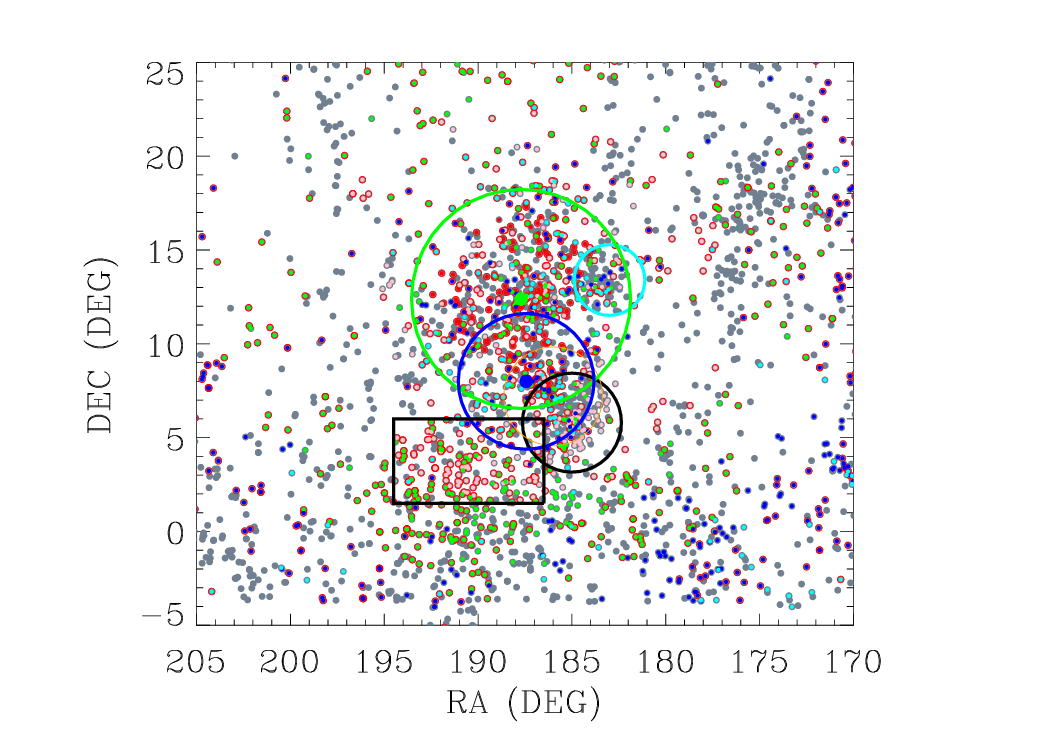}
\caption{Sky projected positions of galaxies in the Virgo cluster region. The locations of M87 and M49 are shown with large green and blue dots, respectively, whereas the large green, blue, black and cyan circles represent the virial radii of M87, M49, W, and M clouds, respectively \citep{Boselli2018}. The black square indicates the position of the W plume. Colors indicate the 3D cluster-centric distance from M87: $r/R_{200} < 1$ (red), $1 < r/R_{200} < 2$ (pink), $2 < r/R_{200} < 3$ (green), $3 < r/R_{200} < 4$ (cyan), and $4 < r/R_{200} < 5$ (blue), while gray points represent galaxies at $r/R_{200} > 5$. We note that the figure shows projected sky positions, whereas the color-coding is based on the three-dimensional distance, so galaxies at large distances can appear close to the cluster in projection (see text for further details).}
\label{fig:virgo}
\end{figure}

The Virgo Cluster, the nearest massive galaxy cluster at $\sim16.5$~Mpc \citep{Mei2007}, is still assembling. Its galaxy distribution, radial velocities, and X-ray morphology reveal multiple accreting substructures \citep{deVaucouleurs1961, Tully1984, Binggeli1993, Mei2007, Kim2016, Boselli2018, Castignani2022a}. Two main subclusters dominate: SubA, centered on M87 ($V_{\rm helio} = 955$~km~s$^{-1}$), and SubB, centered on M49  \citep[$V_{\rm helio} = 1134$~km~s$^{-1}$;][]{Binggeli1993, Boselli2018}.  At larger distances, several galaxy groups—originally identified as clouds by \cite{deVaucouleurs1961}—trace the cluster outskirts. These include the W, W$^{'}$, and M clouds, each with typical group masses $M_{200} \sim 10^{13}$~M$_{\odot}$ and sizes $R_{200} \sim 0.5-0.8$~Mpc \citep{Boselli2018}. The W ($V_{\rm helio} = 2176$ km s$^{-1}$, \cite{Boselli2018}) and M \citep[$V_{\rm helio} = 2109$ km s$^{-1}$,][]{Boselli2018} clouds are behind Virgo, whereas W$'$ has velocities similar to Virgo members  \citep[$V_{\rm helio} = 1019$ km s$^{-1}$,][]{Boselli2018, Kashibadze2020}. Figure \ref{fig:virgo} shows their projected sky distribution.

Among these, the W cloud has been the subject of a long-standing debate in the literature. Its large projected separation from the cluster center and its high relative velocity have led to different interpretations, ranging from a background structure unrelated to Virgo to an infalling or interacting group in the cluster outskirts \citep{deVaucouleurs1961, Tully1984, Binggeli1993, Boselli2018, Cantiello2024}. It has also been suggested that the W cloud may correspond to a filament viewed nearly along its axis \citep{Tully1982}. This ambiguity makes the W cloud a particularly interesting laboratory for studying the ongoing assembly of the Virgo cluster.

\subsection{The Virgo infall cluster catalog (VICC)}

We constructed a parent sample of objects classified as GALAXY in the SDSS-DR16 photometric catalog and located within the Virgo infall region, defined by $100 \leq \mathrm{RA} \leq 280$ and $-5^{o} \leq \mathrm{DEC} \leq 75^{o}$. We included galaxies with spectroscopic redshifts available in both SDSS-DR16 \citep[][]{Ahumada2020} and DESI-DR1 \citep[][]{DES2016}. By applying a velocity cut of $cz \leq 4000$ km s$^{-1}$, we obtained a final sample of approximately 43,975 objects\footnote{The full catalog and the selection of the targets will be presented in detail in a separate work in preparation}.

Although our parent catalog consists of objects classified as GALAXY in SDSS-DR16, the dataset is not free from stellar contamination. This is particularly relevant in the redshift range of the Virgo cluster, where faint stars with $|cz| \le 500$~km~s$^{-1}$ can be misclassified as galaxies. To clean our selection from stars, we ran our own galaxy and star classification, which was carried out using a supervised machine learning approach based on the k-nearest neighbors (KNN) algorithm \citep[][]{Fix1951, Hastie2009, Bishop2006}. KNN is a nonparametric method that assigns the class of a given object according to the majority vote of its k closest neighbors in a multidimensional feature space. Distances are typically computed in Euclidean space after normalizing each parameter, ensuring that all observables contribute equally to the classification. The algorithm has the advantage of being simple, robust, and well-suited for problems where the boundary between classes is nonlinear and must be empirically determined from the data.

In our case, the input feature space included photometric and structural parameters such as the apparent $g-$ and $r$-band magnitudes ($m_{g}$, $m_{r}$), color index ($m_{g}-m_{r}$), mean effective surface brightness ($\mu_e$), and light concentration indices derived from the radii enclosing 50$\%$ and 90$\%$ of the total flux ($r_{50}$, $r_{90}$, and $C=r_{90}/r_{50}$). We have also included the heliocentric velocity of the objects ($v_{helio} = cz_{helio}$) in the multidimensional space. A labeled training set, composed of objects with secure spectroscopic classification, was used to calibrate the KNN classifier. The training set used for the KNN classifier was constructed from objects with reliable spectroscopic classifications available in the DESI and the NED Local Volume Sample \citep[NED-LVS;][]{Cook2014,Cook2023} catalogs. In the DESI catalog, objects are divided into three main spectral classes: STAR, GALAXY, and UNKNOWN. We adopted this classification directly as input for our KNN analysis. Additionally, we considered as GALAXY all objects from the NED-LVS catalog with available redshift-independent distance estimates, since these measurements provide a robust confirmation of their extragalactic nature. Combining these two sources, our training set comprises a total of 17,363) stars and galaxies with spectroscopic classifications, representing about 37$\%$ of the full sample. 

Cross-validation was performed to optimize the number of neighbors and to assess the performance of the separation. The final model was then applied to the full sample, providing a probability-based classification into stars and galaxies, and yielding a confusion matrix that quantifies the robustness of the method.

The performance of our KNN classifier was evaluated using cross-validation on the spectroscopically classified training set. The results show an overall accuracy of 0.98, indicating that 98$\%$ of the objects were correctly identified as either stars or galaxies. For the galaxy class, we obtained a precision of 0.99, meaning that almost all objects classified as galaxies are true galaxies, with very few stellar contaminants. The recall for galaxies is  0.96, showing that the vast majority of true galaxies were correctly recovered by the classifier, with only a small fraction misclassified as stars. These metrics confirm that our classification scheme is both highly reliable and efficient for distinguishing galaxies from stars in our sample.

We applied the KNN classifier to the full sample of objects, obtaining 9,872 galaxies and 34,103 stars. A small fraction of the sources classified as stars (about 1.1$\%$) are in fact not genuine stars, as they show heliocentric velocities $v_{\rm helio} > 500$ km s$^{-1}$ or $v_{\rm helio} < -250$ km s$^{-1}$. This misclassification arises because only a few galaxies in our training set have $v_{\rm helio} < 0$, which causes their k nearest neighbors to be predominantly classified as stars. This issue reflects the nonhomogeneous distribution of objects in the adopted multiparameter space, where some sources are more isolated than others. Since KNN assigns classes based on the majority vote of the k nearest neighbors, the classification of such isolated objects is intrinsically less reliable. To mitigate this effect, we computed a reliability flag defined as the minimum distance ($d_{\rm min}$) from each object to its k nearest neighbors. We then defined a secure sample of galaxies as those with $d_{\rm min} < d_{95}$, where $d_{95}$ corresponds to the 95th percentile of the $d_{\rm min}$ distribution for galaxies. This selection reduce the misclassified stars to about $0.7 \%$. Following this approach, we obtain a final secure sample of 9,379 galaxies.

Finally, we cross–matched our Virgo infall catalog with the EVCC and the Castignani catalogs \citep{Kim2014,Castignani2022a}, yielding a final sample of 11,097 secure galaxies with spectroscopic redshifts across the full Virgo infall region and homogeneus photometric data from SDSS-DR16. This new sample is twice as large as previous catalogs \citep[][]{Castignani2022a}.

\subsection{Cosmic distances}

The position of the different galaxies within the infall cluster region of Virgo have been obtained by their intrinsic distances. We have achived this result by doing several corrections to the heliocentric redshift measurements provided by our parent catalog.

To place our galaxies in a consistent reference frame and facilitate comparison with other studies, we first convert the measured heliocentric velocities to the LG frame. This transformation accounts for the Sun’s motion relative to the LG barycenter and follows

\begin{equation}
v_{\rm LG} = v_{\rm helio} + V_\odot \left[\sin(b)\sin(b_\odot) + \cos(b)\cos(b_\odot)\cos(l-l_\odot)\right],
\end{equation}

where $v_{\rm helio}$ is the heliocentric velocity, $(l, b)$ are the Galactic coordinates of the galaxy, and $V_\odot \approx 316$ km~s$^{-1}$ is directed toward $(l_\odot, b_\odot) = (93^\circ, -4^\circ)$ \citep{Yahil1977, Karachentsev1996}. This correction is particularly relevant for nearby galaxies, where peculiar motions can be comparable to the Hubble expansion.

Beyond the LG transformation, we must also account for the peculiar velocity field induced by the Virgo cluster. The gravitational pull of Virgo causes significant deviations from a pure Hubble flow, called the Virgo infall velocity. To achive this we modeled Virgo as a spherical overdensity with a Navarro–Frenk–White (NFW) profile \citep{Navarro1997}, centered at the cluster (M87) position . The density distribution is

\begin{equation}
\rho(r) = \frac{\rho_s}{(r/r_s)(1+r/r_s)^2},
\end{equation}
with scale radius $r_s = R_{200}/c$ and characteristic density

\begin{equation}
\rho_s = \frac{200}{3}\,\rho_c \,\frac{c^3}{\ln(1+c)-c/(1+c)},
\end{equation}
where $\rho_c = 3H_0^2/(8\pi G)$ is the critical density of the Universe. The enclosed mass and mean overdensity within radius $r$ are given by:

\begin{equation}
M(<r) = 4\pi\rho_s r_s^3\left[\ln\left(1+\frac{r}{r_s}\right)-\frac{r}{r+r_s}\right],
\end{equation}
\begin{equation}
\bar{\delta}(<r) = \frac{3M(<r)}{4\pi r^3\rho_b} - 1,
\end{equation}
with $\rho_b=\Omega_m\rho_c$. The radial infall velocity can be estimated using linear theory \citep{Peebles1980},

\begin{equation}
v_{\rm lin}(r) = -\frac{1}{3} H_0 \Omega_m^{0.6}\, r\, \bar{\delta}(<r),
\end{equation}
or the nonlinear Yahil--type correction taken from the implementation given by \citet{Croft1999}.

\begin{equation}
v_{\rm nonlin}(r) = -\frac{1}{3} H_0 \Omega_m^{0.6}\, r\, \bar{\delta}(<r)\,\left[1+\bar{\delta}(<r)\right]^{-1/4}.
\end{equation}

We adopted the nonlinear approach resulting the observable line-of-sight component as

\begin{equation}
v_{\rm infall} = v_{\rm nonlin}(r)\,\cos\theta,
\end{equation}
where $\theta$ is the angle between the radial vector from Virgo and the line of sight. For the Virgo cluster we adopted $M_{200} =  4.2 \times10^{14} M_{\odot}$, $R_{200} = 1.55$ Mpc, and $c=2.8$ \citep[][]{McLaughlin1999}.

The final cosmic velocity of each galaxy is therefore 

\begin{equation}
v_{\rm cosmic} = v_{\rm LG} + v_{\rm infall},
\end{equation}
and the corresponding distance is $d_{\rm cosmic} = v_{\rm cosmic}/H_{0}$. 

\begin{figure}
\centering
\includegraphics[width=0.5\textwidth]{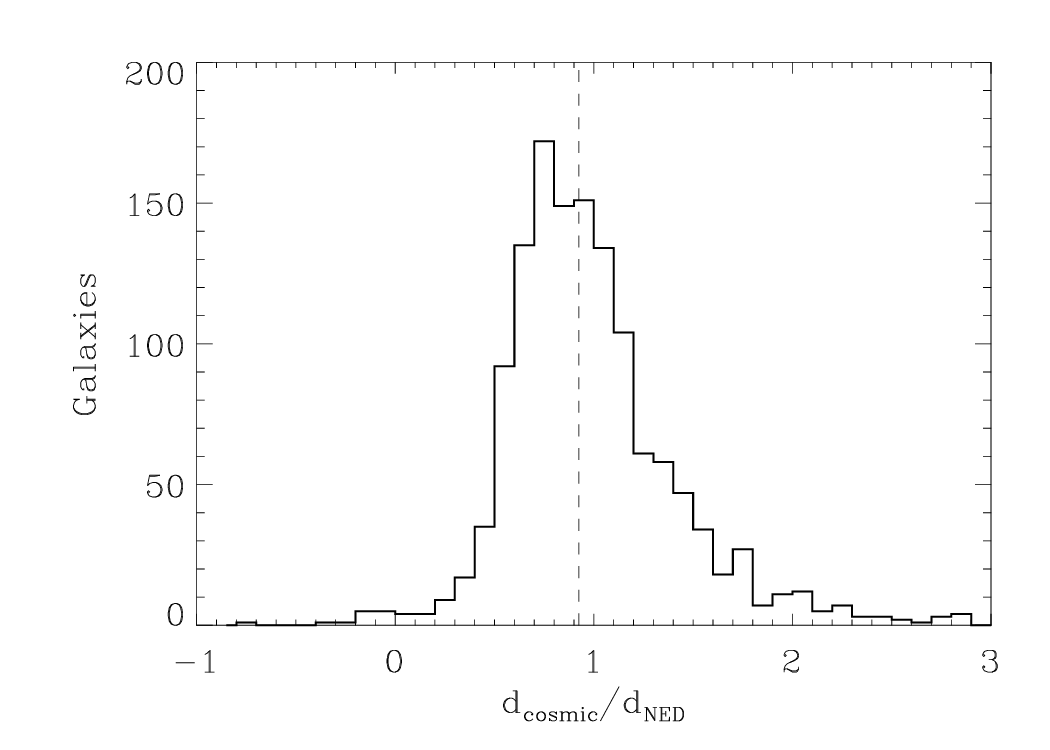}
\caption{Ratio of our computed cosmic distances to those obtained by z-independent methods in NED-LVS. The vertical dashed line represent the median value of the ratio $d_{cosmic}/d_{NED}$.}
\label{fig:dcosmic-dned}
\end{figure}

To evaluate the accuracy of our infall-corrected distances, we have also incorporated the NED-LVS catalog, which contains information on about 2 million objects out to distances of $\sim$1000 Mpc \citep[][]{Cook2023}. For objects within 200 Mpc, this catalog provides redshift-independent distance estimates, which are particularly valuable for identifying galaxies belonging to the Virgo cluster and its surrounding filaments. Cross-matching our parent catalog with NED-LVS yields a total of 7,656 common objects, among which 1,519 have redshift-independent distance measurements. We compared $d_{\rm cosmic}$ with the redshift–independent measurements available in the NED–LVS database ($d_{\rm NED}$). Figure~\ref{fig:dcosmic-dned} shows the distribution of the ratio $d_{\rm cosmic}/d_{\rm NED}$. For our model, the median and scatter of this ratio are 0.92 and 0.43 , respectively. 

These values represent an improvement over the standard Mould correction \citep{Mould2000}, which yields a median of 0.89 and a scatter of 0.39 for the same comparison.

\subsection{Supergalactic coordinates}

To place our galaxies in the context of the local large-scale structure, we transformed their positions from Galactic coordinates and distances to the supergalactic Cartesian system \citep{deVaucouleurs1953, Lahav2000}. In this frame, positions are expressed as $(SGX, SGY, SGZ)$, where the origin corresponds to the LG and the plane of the Local Supercluster roughly coincides with $SGZ = 0$.

The transformation proceeds in two steps. First, we convert the right ascension and declination of each galaxy to Galactic coordinates $(l, b)$, and apply the rotation matrix to align the Galactic system with the supergalactic system. Second, we combine the resulting supergalactic longitude and latitude $(SGL, SGB)$ with the cosmic distance $d_{\rm cosmic}$ of each galaxy, to compute the Cartesian components (SGX, SGY, SGZ). In this coordinate system the Virgo cluster is located at SGX$\sim -3$ Mpc, SGY $\sim $16. Mpc, SGZ $\sim - 0.7$ Mpc \citep[see][]{deVaucouleurs1976, Tully2013}.

This three-dimensional representation allows us to study the spatial distribution of our sample relative to the main structures of the Local Supercluster, such as the Virgo cluster and nearby filaments. Using $(SGX, SGY, SGZ)$ coordinates is particularly useful for visualizing alignments with the cosmic web and for comparing peculiar velocities and infall patterns along preferred directions in the supergalactic plane.

Figure~\ref{fig:virgo} shows the projected sky distribution of galaxies in the Virgo cluster region within $5 \times R_{200}$ of M87. Galaxies are color-coded according to their three-dimensional distance from M87, computed using the supergalactic coordinates described above. Because the figure shows projected positions, galaxies located at large three-dimensional distances may appear close to the cluster in the sky. The figure highlights the complex spatial structure of the Virgo environment, including the main subclusters and the surrounding outskirts region. This galaxy sample forms the basis of the analysis presented in this paper.

\subsection{Selection of the W cloud and W plume}

The selection of galaxies associated with the W cloud was initially based on its projected position and characteristic radius, $R_{200}$, as reported by \cite{Boselli2018}. We first selected 436 galaxies within a projected radius of 0.8~Mpc centered on $(\mathrm{RA},\mathrm{DEC})=(185.00^\circ,5.80^\circ)$.

\begin{figure}
\centering
\includegraphics[width=0.4\textwidth]{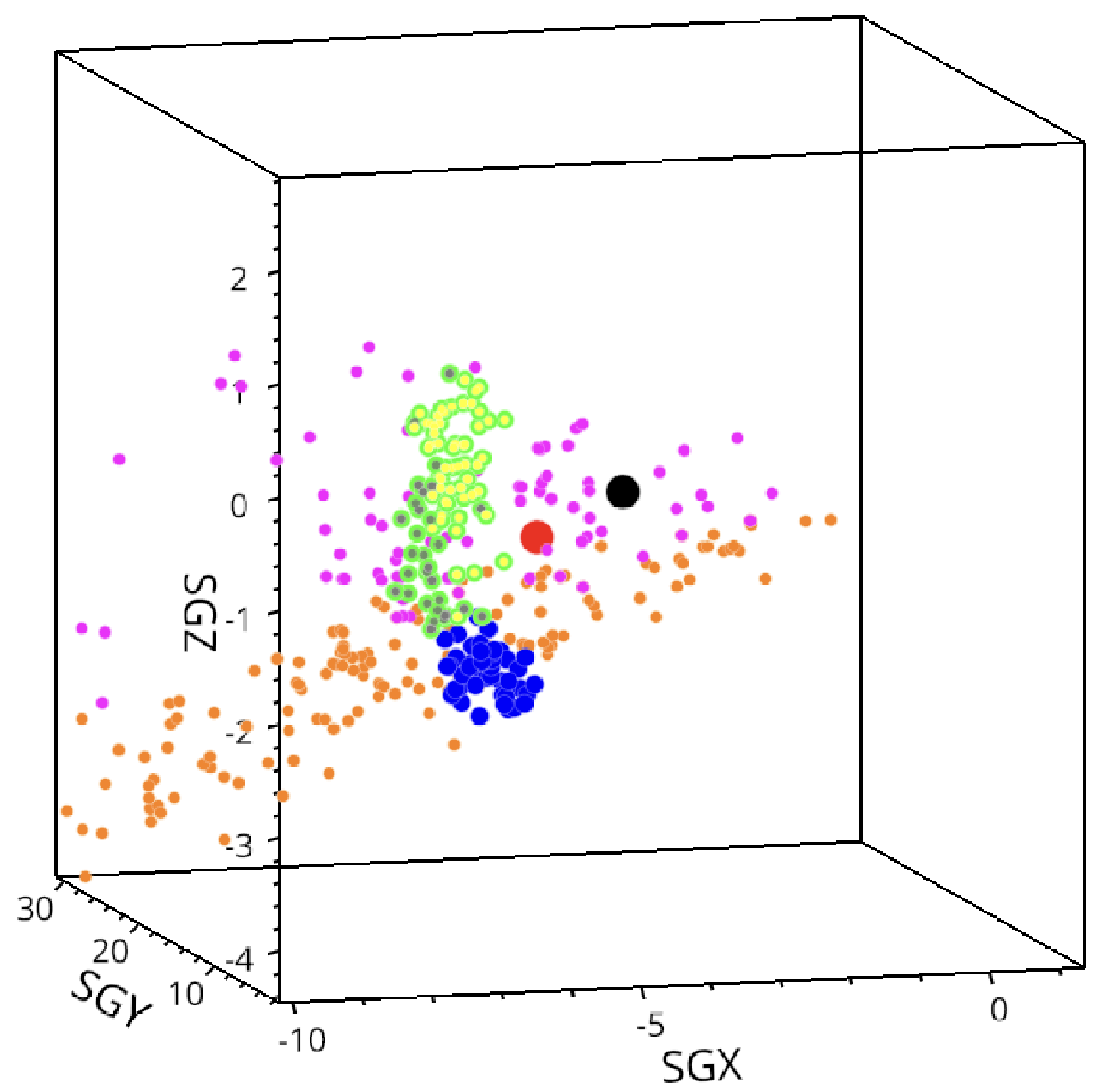}
\caption{Three–dimensional view of the W cloud (orange) and W plume (magenta). The positions of M87 and M49 are shown as black and red points. The W group and the W tail are highlighted in blue and green, respectively. The yellow and gray symbols overlaid on the green points correspond to W tail galaxies with $V_{\rm LG}<1200$ and $V_{\rm LG}>1200$~km~s$^{-1}$ (see text for more details). }
\label{fig:3dview}
\end{figure}

To characterize the three-dimensional structure of this sample, we made use of the supergalactic coordinates of the selected galaxies. Their resulting three-dimensional distribution, shown in Fig.~\ref{fig:3dview} with orange and blue symbols, is globally elongated, suggestive of a filamentary structure, and contains a prominent overdensity in phase space located at $(\mathrm{SGX},\mathrm{SGY},\mathrm{SGZ}) =  (-5.58,\,16.63,\,-2.23)$. This overdensity is clearly distinguished from the surrounding galaxy population. We identify this structure as the W group (blue points in Fig.~\ref{fig:3dview}). It comprises 50 galaxies located within a three-dimensional radius of 0.8~Mpc from the overdensity center, providing a physically motivated definition of the group based on its full spatial distribution rather than on projected properties alone. The 50 group galaxies show a mean velocity of $V_{\rm LG}=2140$~km~s$^{-1}$ and a velocity dispersion of $\sigma=140$~km~s$^{-1}$. From the supergalactic coordinates of the W group we derive a distance of 17.7 Mpc from the Milky Way. The three-dimensional separation between the Virgo cluster center and the W group is about 2.5 Mpc, indicating that the group lies behind the Virgo cluster along the line of sight.

We initially define the W plume as the population of galaxies located within the rectangular sky region 
$186^\circ \lesssim \mathrm{RA} \lesssim 195^\circ$ and $1^\circ \lesssim \mathrm{DEC} \lesssim 6^\circ$. 
This area encompasses a high fraction of galaxies located at cluster-centric distances 
$1 < r/R_{200} < 2$ from M87 (pink symbols in Fig.~\ref{fig:virgo}), corresponding to the outer regions of the Virgo cluster. This region of the sky broadly overlaps with the east and west Southern Extensions defined by \cite{Kim2014}. The full W plume sample contains a total of 213 galaxies.

To investigate the three-dimensional structure of this population, we examined the distribution of these galaxies in supergalactic coordinates. The resulting 3D configuration is shown in Fig.~\ref{fig:3dview} (magenta and green symbols). In contrast to the W cloud, the three-dimensional distribution of galaxies in the W plume is visually less elongated. Instead, the W plume galaxies occupy a coherent region of supergalactic space, characterized by mean coordinates $(\mathrm{SGX},\mathrm{SGY},\mathrm{SGZ}) =(-7.75,\,19.12,\,-0.64)$. These coordinates differ from the SE reported by \cite{Tully1982}. In particular, the SE spans a sky region of $180^{\ o} < RA < 192^{\ o}$ and $-20^{\ o} < DEC < +5^{\ o}$ \citep[see][]{Karachentsev2013}. Our W plume region coincides with northern part of this larger area. However, most of the galaxies located in the W plume are found within a narrow 3D distance region of $1 < r/R_{200} < 2$ in contrast with other galaxies in the rest of the SE, which are found at larger 3D distances.

We identify a particularly prominent overdensity of galaxies of the W plume in three-dimensional space, centered at 
$(\mathrm{SGX},\mathrm{SGY},\mathrm{SGZ}) \simeq (-6.5,\,15.0,\,-1.25)$. 
This region occupies a volume approximately bounded by 
$\Delta\mathrm{SGX}\simeq2$~Mpc, 
$\Delta\mathrm{SGY}\simeq2$~Mpc, and 
$\Delta\mathrm{SGZ}\simeq3$~Mpc, 
and contains 94 galaxies. Its elongated three-dimensional morphology resembles a tail-like structure; we therefore define this dense substructure as the W tail (green points in Fig.~\ref{fig:3dview}), which represents the core of the plume in three-dimensional space.

These definitions rely exclusively on the spatial distribution of galaxies on the sky and in supergalactic coordinates, and therefore provide a purely geometrical foundation for the subsequent statistical and kinematic analysis. Within this framework, we aim to demonstrate that the W group constitutes the core of the W cloud, while the W tail represents a spatially and dynamically coherent population of galaxies that is being stripped from the W group as a result of its interaction with the cluster potential.

\section{Results}

In this section, we present the main results of our analysis of the newly discovered W group and its associated W tail. We first examine the velocity structure of the system, highlighting the kinematic connection between the group and the extended tail. We then explore the three-dimensional configuration of the system, combining positional and velocity information to constrain its spatial extent and dynamical state. Finally, we analyze the stellar properties of galaxies in the W group and the W tail, providing insight into the environmental processes shaping galaxy evolution during group accretion.

\subsection{Velocity structure}

\begin{figure}
\centering
\includegraphics[width=0.5\textwidth, trim = 10 20 10 30]{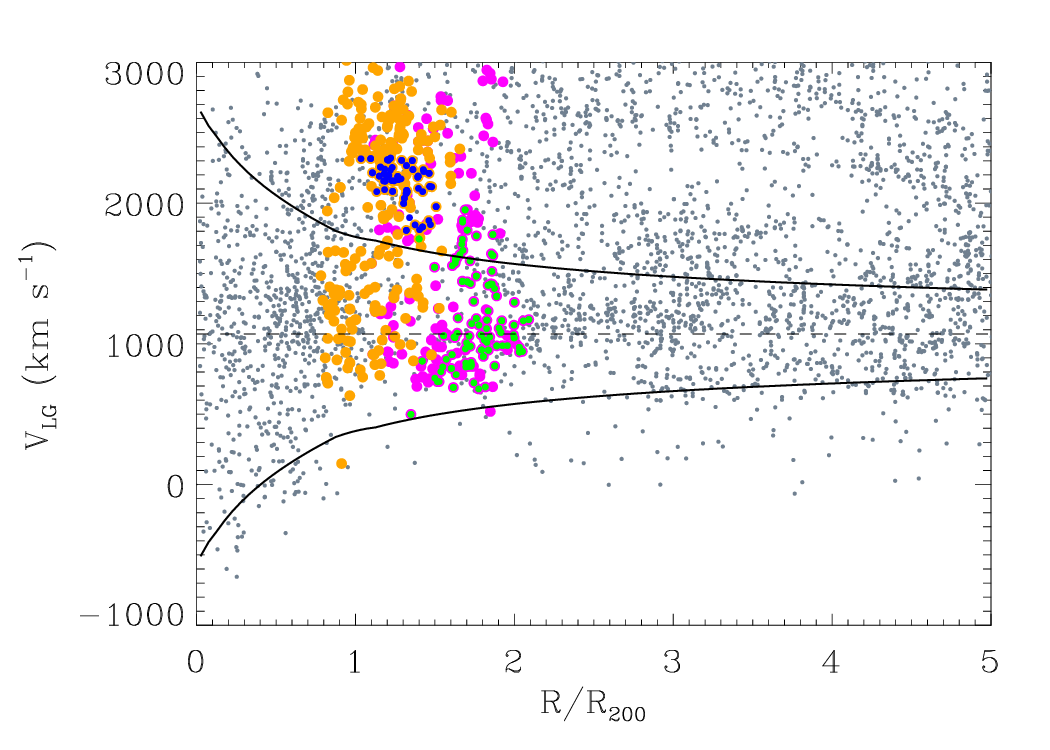}
\caption{Sky projected phase-space diagram of galaxies within $R/R_{200} < 5$ in the Virgo cluster region. Colors indicate the W cloud (orange), W  group (blue), W plume (magenta) and W tail (green; see text for details about the selection of the different structures of the W cloud and W plume). M87 was considered as center of the Virgo cluster.}
\label{fig:caustica_virgo}
\end{figure}

The projected phase-space distribution of Virgo's galaxies is shown in Figure \ref{fig:caustica_virgo}, with the caustic overplotted on it. Most of the W cloud galaxies (orange points) lie outside the caustic -- they are not dynamically bound to Virgo -- while the W plume galaxies (magenta points) are more equally distributed inside and outside the Virgo caustic.  In the caustic diagram, galaxies from the W group (blue points) and the W tail (green points) are overplotted. The W group galaxies lie outside the Virgo caustics, indicating that the group as a whole is not gravitationally bound to the cluster. In contrast, most of the W tail galaxies lie within the caustics and are therefore already bound to the cluster, although a small fraction remains unbound.

\begin{figure}
\centering
\includegraphics[width=0.5\textwidth]{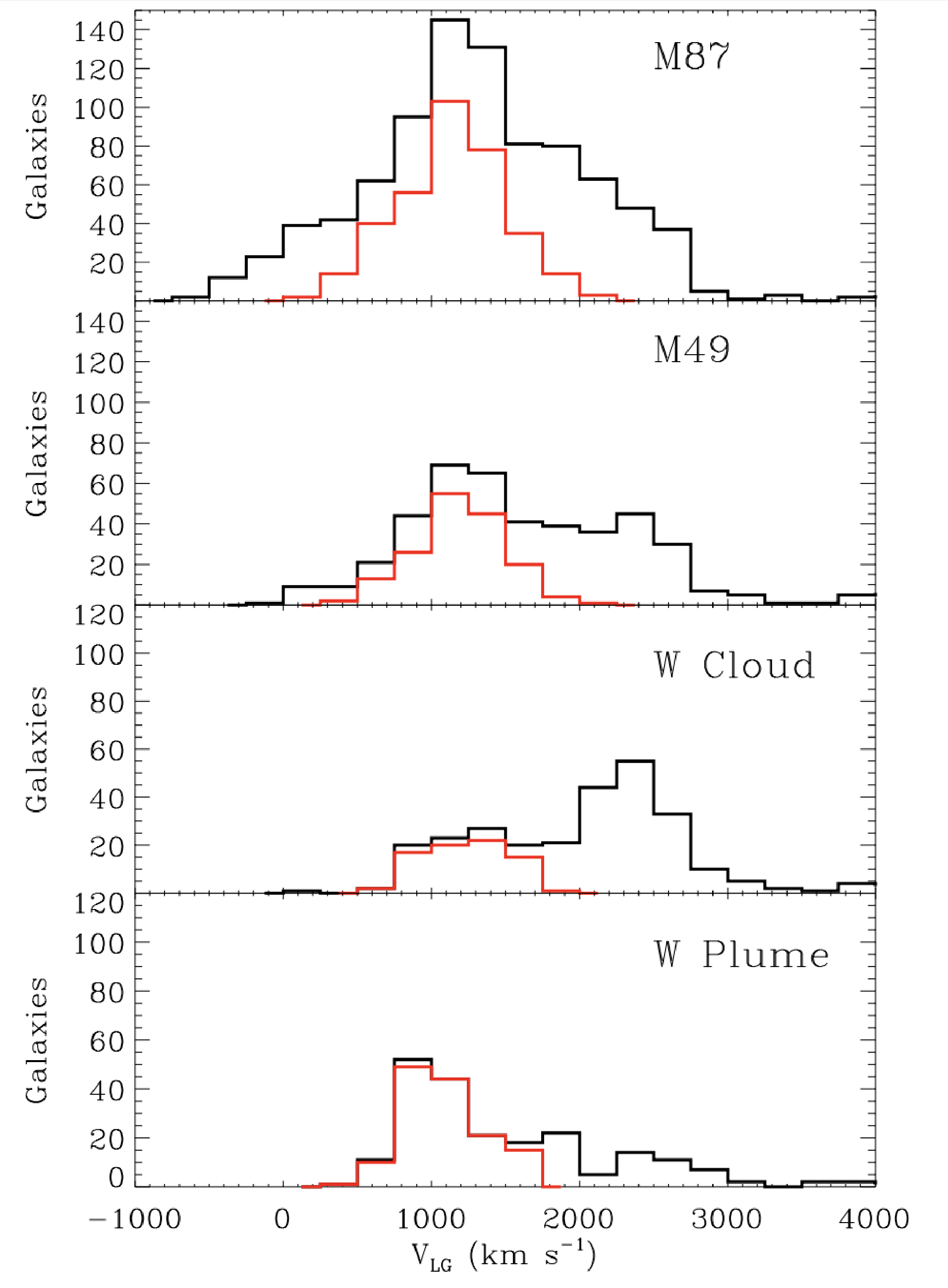}
\caption{Velocity distributions of the galaxies in the M87, M49, W cloud, and W plume regions. The black histograms correspond to all galaxies at those locations and the red histograms to those galaxies located within the Virgo caustic. The galaxies of the different regions were selected as in Fig. \ref{fig:virgo}.}
\label{fig:sub_vel}
\end{figure}

Figure~\ref{fig:sub_vel} shows the velocity distributions of the main Virgo substructures. The velocity components of the W group and the W tail identified in the caustic diagram are also clearly reflected in this figure. The W cloud exhibits a bimodal velocity distribution, with peaks at low ($\sim1200$ km~s$^{-1}$) and high ($\sim2400$ km~s$^{-1}$) velocities (bottom left panel in Fig. \ref{fig:sub_vel}). The low-velocity peak corresponds to galaxies that are gravitationally bound to the Virgo cluster, whereas the high-velocity peak is predominantly composed of W group galaxies and lies outside the cluster caustics. In contrast, the W plume is dominated by a low-velocity component ($\sim1000$ km~s$^{-1}$) whose galaxies are largely gravitationally bound to the cluster, while also displaying an extended tail toward higher velocities. The W tail galaxies predominantly belong to the low-velocity peak (red line in bottom left panel of Fig. \ref{fig:sub_vel}), although a fraction of them exhibit higher velocities.

\begin{figure}
\centering
\includegraphics[width=0.5\textwidth, trim = 80 10 125 20]{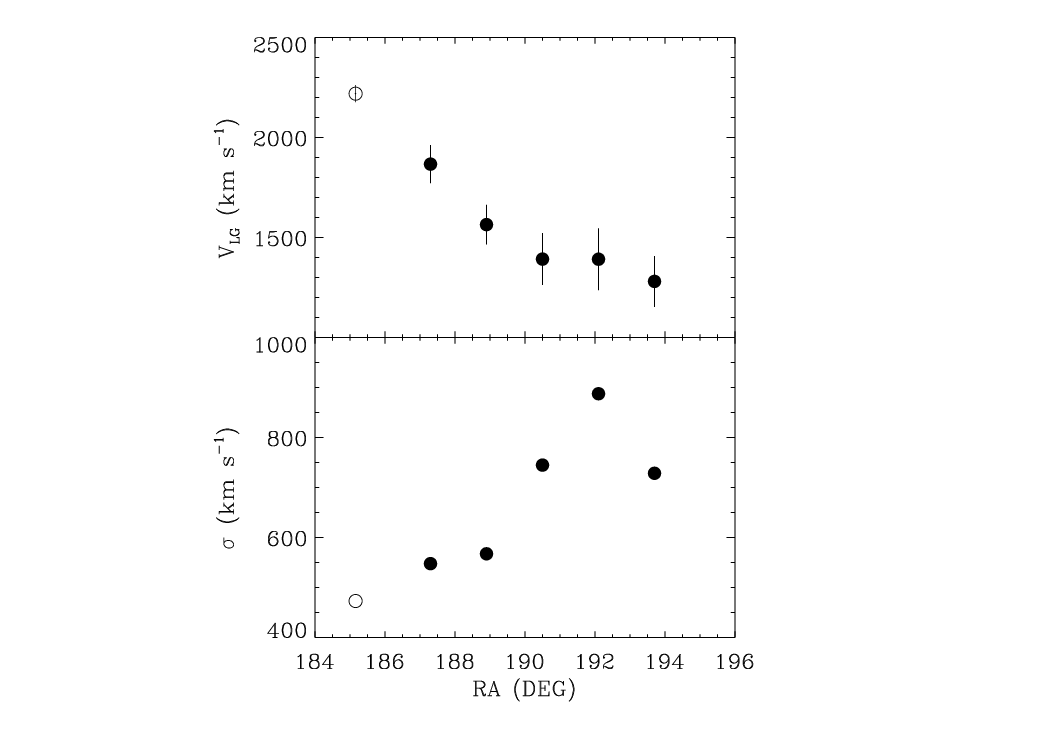}
\caption{Radial velocity (top panel) and dispersion velocity (bottom panel) of galaxies in the W plume located at different RA locations (distance to the W cloud). The empty circle represents the radial velocity and dispersion of the galaxies in the high-velocity tail of the W cloud ($V_{LG} > 1800$ km s$^{-1}$).}
\label{fig:vel_sig_WPlume}
\end{figure}

To quantify the dynamical connection between the W cloud and the W plume, we first analyzed the kinematics of W plume galaxies in projection. We measured the mean line-of-sight velocity and velocity dispersion of W plume galaxies as a function of projected right ascension distance from the center of the W cloud (Fig.~\ref{fig:vel_sig_WPlume}). We find a systematic decrease in the mean velocity accompanied by a clear increase in the velocity dispersion toward the Virgo cluster. These coherent kinematic gradients demonstrate that the W plume is not a dynamically independent structure, but is closely linked to the high-velocity component of the W cloud identified in the projected phase space. Such trends are characteristic of material tidally stripped from an infalling galaxy group and progressively responding to the deeper gravitational potential of the cluster \citep[e.g.,][]{Mihos2005,Rudick2009,Vijayaraghavan2015,Rhee2017}.

While this connection was initially identified in projection, it becomes clearer when the system is examined in three-dimensional space. The velocity gradient observed in projection is preserved in 3D (Fig.~\ref{fig:3dview}), ruling out projection effects as its origin. In three dimensions, the dynamical link is revealed to be between the W group and the W tail. In particular, W tail galaxies located at smaller three-dimensional distances from the W group exhibit systematically higher velocities (gray symbols in Fig.~\ref{fig:3dview}) than those located at larger 3D distances (yellow symbols). This continuous variation of velocity with spatial separation defines a clear dynamical bridge between the W group and the W tail, consistent with a tidal origin.

The combined kinematic and spatial information further constrains the interaction sequence. W tail galaxies with lower line-of-sight velocities and located at larger three-dimensional distances from the W group are associated with the prominent low-velocity peak in the velocity distribution (Fig.~\ref{fig:sub_vel}). These galaxies lie within the Virgo cluster caustics (Fig.~\ref{fig:caustica_virgo}) and are therefore gravitationally bound to the cluster.

In contrast, galaxies associated with the high-velocity tail of the W plume velocity distribution are predominantly found at smaller 3D distances from the W group and lie outside the Virgo caustics, indicating that they are not yet gravitationally bound to the cluster. This clear segregation in velocity and phase space naturally defines an evolutionary sequence: galaxies stripped earlier during the group--cluster interaction have lost orbital energy and are now bound to Virgo, while those stripped more recently remain dynamically connected to the W group and are still in the process of infall.

\subsection{Three-dimensional structure}

Figure~\ref{fig:3dview} shows that the three-dimensional distribution of galaxies in the W cloud is globally elongated, suggestive of a filamentary structure within which the W group is embedded. Such an interpretation was already proposed by \cite{Tully1982} to explain the nature of the W cloud. If the W cloud traces a filament of the large-scale structure, the dynamical interpretation of the interaction between the W group and the Virgo cluster would be different, as the group could be infalling into the cluster along this filament.

Importantly, the connection between the W group and the W tail is not limited to a projected alignment on the sky. When analyzed in three-dimensional space, the W tail remains spatially and kinematically connected to the W group, defining a continuous structure in both position and velocity. This three-dimensional continuity strongly favors a physical association between the two components and indicates that the W tail does not simply trace the large-scale filamentary distribution of galaxies, but instead represents material dynamically linked to the W group and recently stripped during its interaction with the Virgo cluster.

To determine whether the W cloud is a genuine filament rather than a projected configuration, we quantitatively analyzed the three-dimensional geometry of its galaxy distribution. In particular, we tested whether the W cloud could correspond to a filamentary structure observed close to its axis by computing the eigenvalues of its 3D spatial distribution.

We computed the shape tensor of the galaxy distribution using the covariance matrix of the centered Cartesian coordinates (SGX, SGY, SGZ). For each region, we derived the three eigenvalues and eigenvectors of the $3\times 3$ covariance matrix. The principal axis of the structure was defined as the eigenvector associated with the largest eigenvalue, and the filament elongation (E) was quantified as

\begin{equation}
{\rm E} \;=\; \sqrt{\frac{\lambda_{1}}{(\lambda_{2}+\lambda_{3})/2}}\,,
\end{equation}
where $\lambda_{1}\ge\lambda_{2}\ge\lambda_{3}$ are the ordered eigenvalues. The projection of each galaxy onto the principal axis was used to derive the coordinate $s$ along the axis, and the perpendicular distance $r_{perp}$ was computed from the residuals of the projection. From these distances we constructed the transverse density profile of the galaxy distribution around the axis.

\begin{figure*}
%\centering
\sidecaption
\includegraphics[width=0.7\textwidth, trim = 0 40 0 0]{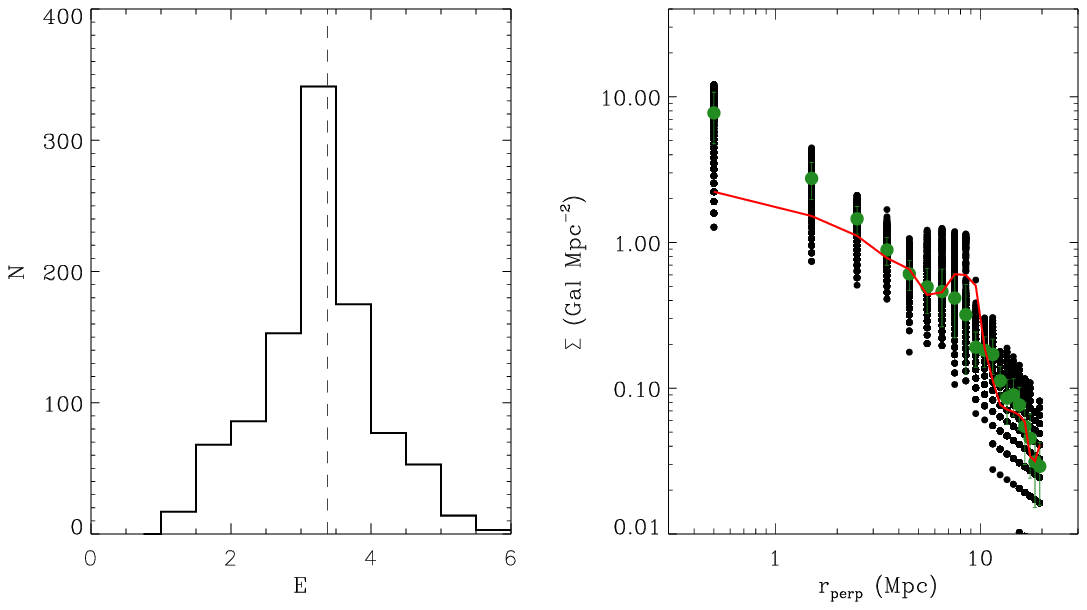}
\caption{(Left) Distribution of elongation values for the 1000 random realizations similar to the W cloud. The vertical line marks the elongation measured for the 3D galaxy distribution of the W cloud. (Right) Transverse galaxy density profiles for the same 1000 random realizations (black points). The green points indicate their mean profile, while the red line shows the observed profile of the W cloud.}
\label{fig:wcloud_filament}
\end{figure*}

To evaluate the statistical significance of the W cloud result and exclude selection effects, we repeated exactly the same PCA procedure for 1000 random positions in the Virgo infall region, selected to have similar number of galaxies as the W cloud. The resulting distribution of elongations and transverse profiles from these random controls was used as the reference baseline (see Fig. \ref{fig:wcloud_filament}).

The W cloud shows a well-defined principal axis and an elongation value compatible with an intrinsically elongated configuration, and its transverse density profile displays a mild central excess with a smooth decline at larger $r_{perp}$. However, when compared to the ensemble of 1000 random control positions in the Virgo infall region, both signals are statistically consistent. For the elongation we computed a Monte Carlo p–value, defined as the fraction of random realizations with elongation equal to or larger than the observed value; this yields $p = 0.40$, indicating that the W cloud elongation is entirely typical within the random sample. Likewise, for the transverse density profile we measured the distance between the observed profile and the mean random profile, and compared this to the distribution of distances from the 1000 random realizations to their own mean; the resulting Monte Carlo p–value is $p = 0.07$, again not significant. Therefore, within the sensitivity of this PCA–based shape analysis, the W cloud signal is not distinguishable from statistical fluctuations expected in the Virgo infall region, i.e.\ this candidate does not pass a significance test relative to matched random controls, and the filament hypothesis for the W cloud can be discarded.  We therefore conclude that the W cloud is not an intrinsic filament, but rather a pencil of galaxies which elongated structure is due to the selection effect of using a circular aperture in a dense environment. This pencil distribution is clearly dominated by a group of galaxies just behind Virgo.

\begin{figure}
\centering
\includegraphics[width=0.4\textwidth]{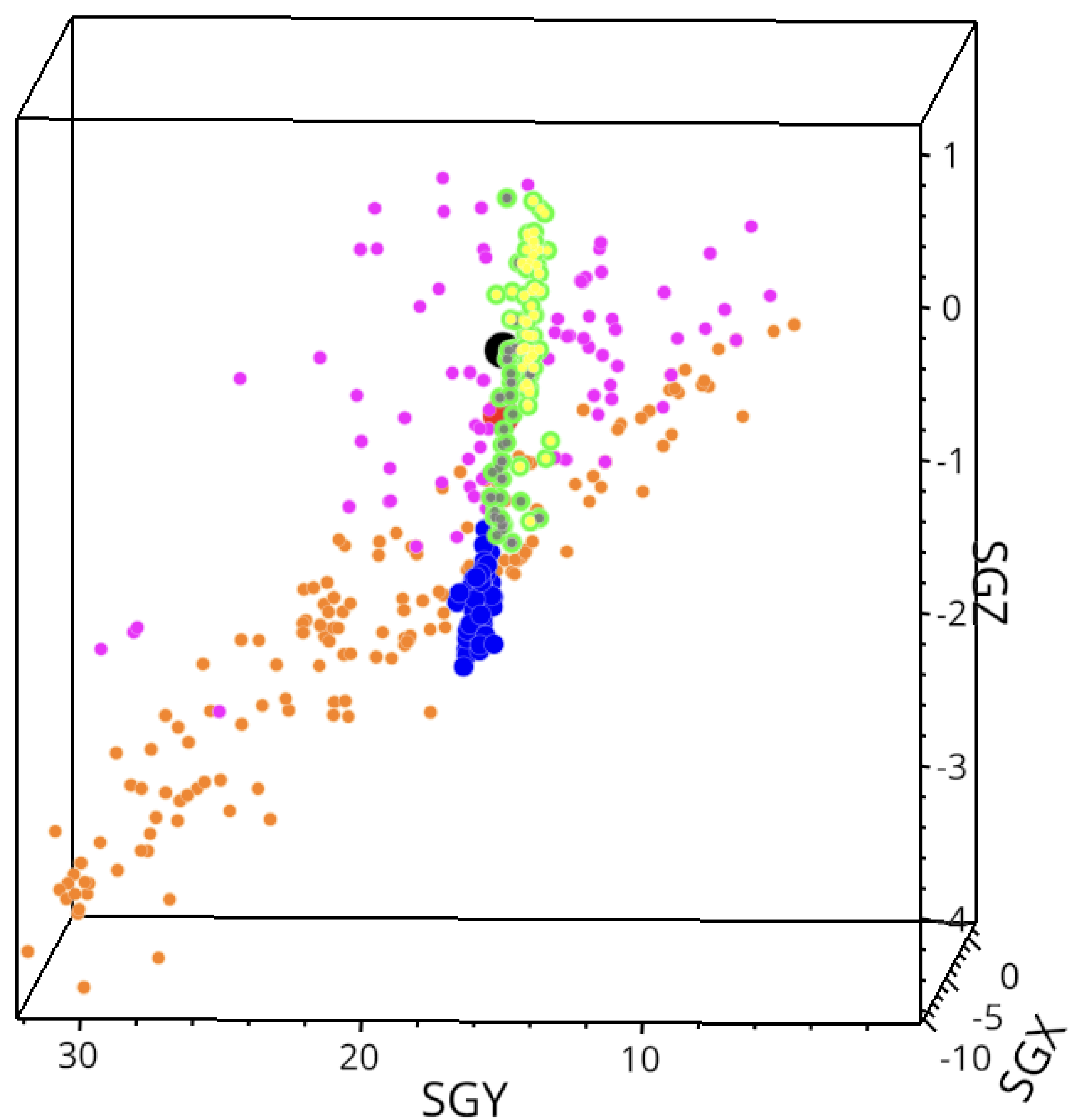}
\caption{Edge–on view of the best–fit orbital plane of the W group–W tail system. The axes are in units of  Mpc and the colors are as in Fig. \ref{fig:3dview}}
\label{fig:3dview-plane}
\end{figure}

We show in Fig. \ref{fig:3dview} that the W plume is significantly more compact than the W cloud. Moreover, it exibits a coherent tail of galaxies toward the Virgo cluster that preserve a clear velocity gradient. It is worth noting that the W tail seems to occupy a well-defined plane configuration, that is clearly visible in a different 3D rotation, similar to the one presented in Fig. \ref{fig:3dview-plane}. We fit this structure with a plane, finding a perpendicular dispersion of $\sigma_{\perp}\simeq0.42$~Mpc and in-plane dispersions of 0.67--0.80~Mpc ($\tau\simeq0.83$). 

Remarkably, the orientation of this plane differs from that of the W group by only $\sim14^{\circ}$ (bootstrap median $=14.1^{+8.4}_{-6.8}$~deg), indicating a shared orbital geometry. This close alignment, together with the spatial continuity and preserved velocity gradient, strongly suggests that the W tail traces a dynamically coherent stream. These properties indicate that the W tail represents tidally stripped material from the infalling W group, dynamically heated during the interaction with the Virgo cluster potential.

\subsection{Galaxy properties}

\begin{figure}
\centering
\includegraphics[width=0.45\textwidth]{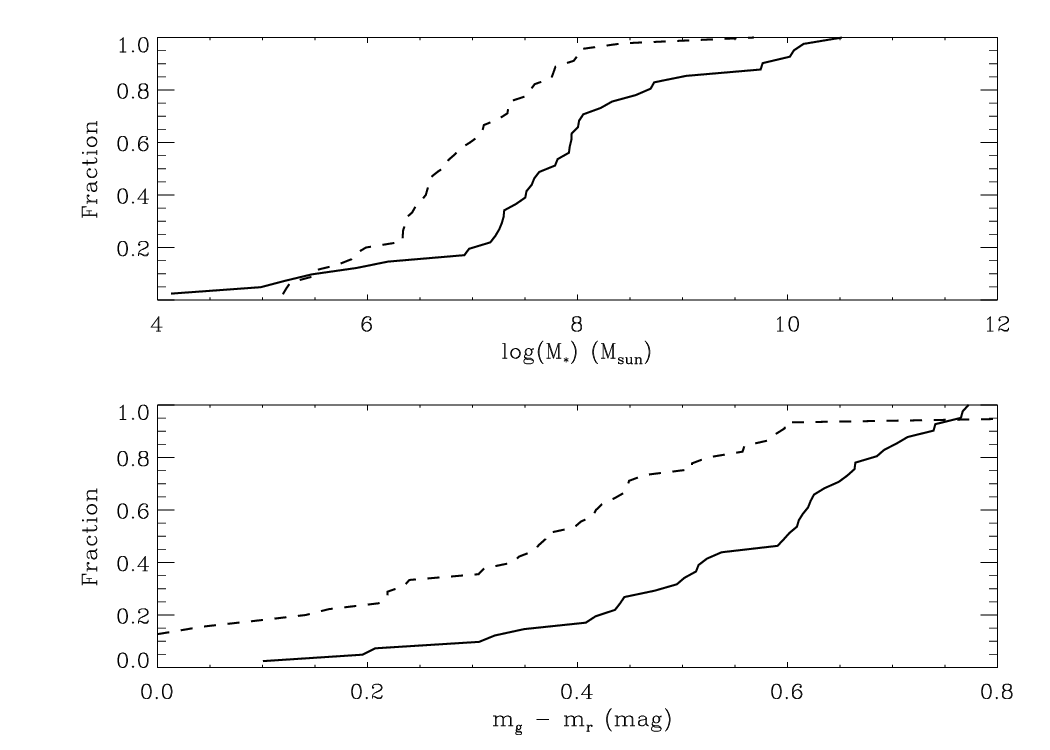}
\caption{(Top panel) Cumulative spectroscopic stellar mass distributions of galaxies of the W group (black solid line) and the W tail (dashed line). (bottom panel) Stellar $g - r$ color for the galaxies of the W group (solid line) and the W tail (dashed line).}
\label{fig:gal_properties}
\end{figure}

Figure \ref{fig:gal_properties} shows the cumulative stellar mass functions of the W group and W tail, with stellar masses derived from FIREFLY/SDSS and CIGALE/DESI spectral fits \citep[][]{Comparat2017,Siudek2024}. These stellar masses are taken directly from the FIREFLY and CIGALE catalogs and were not recomputed in this work. It is worth noting that stellar mass estimates derived from spectral energy distribution fitting are subject to systematic uncertainties, particularly in the dwarf galaxy regime where the presence of young stellar populations can bias the mass-to-light ratio and potentially lead to underestimated stellar masses. \citep[][]{Neumann2022, To2025} The group dynamical mass inferred from $\sigma_{v}=140$ km s$^{-1}$ and $R=0.8$ Mpc is $M_{\rm dyn}\sim10^{13}$ M$_{\odot}$. Its total stellar mass is $M_{\ast}\sim10^{11}$ M$_{\odot}$, yielding $M_{\rm dyn}/M_{\ast}\sim100$, consistent with well-studied dark-matter dominated groups \citep[][]{Karachentsev2002, Mahdavi2005} and abundance-matching expectations \cite{Moster2013}. Within the W group, four galaxies have stellar masses above $10^{10}$~M$_{\odot}$ according to these catalogs (i.e., NGC4255, UGC07422, NGC4269 and VCC260). The most massive galaxy, NGC4255, has $\log(M_{\ast}/M_{\odot}) = 10.52$, in good agreement with other estimates in the literature \citep[e.g.,][]{Galliano2021}.

The stellar mass in the W tail is $M_{\ast,{\rm tail}}\sim10^{10}$ M$_{\odot}$, $\sim10$ percent of the group stellar mass. This indicates partial but significant stripping. Median $(g-r)$ colors are 0.6 (W group) and 0.4 (W tail), indicating that both populations are dominated by blue, star-forming dwarf galaxies and are consistent with systems that have not yet experienced significant environmental quenching. However, the Kolmogorov--Smirnov test applied to the cumulative color distribution of the galaxies in the W group and the W tail indicates that the differences are not statistically significant ($p_{\rm KS} = 0.47$). For stellar masses of $10^7$--$10^8,M_\odot$, galaxies on the red sequence in Virgo typically exhibit $g-r \gtrsim 0.6$--0.7 \citep[e.g.,][]{Janz2009,Janz2021}. The median colors of the W group and especially of the W tail therefore place these galaxies in or near the blue cloud, supporting the presence of young stellar populations. However, the slightly bluer colors observed in the W tail do not necessarily require a specific environmental effect associated with the interaction with the Virgo cluster. Instead, they can naturally arise from the somewhat lower stellar masses of the galaxies in the tail compared to those in the W group, since lower-mass galaxies are expected to have higher specific star formation rates and therefore bluer colors \citep[][]{Janz2009, Janz2021}.

These stellar population properties argue against a backsplash origin for the W tail galaxies. Systems that have already passed through the cluster core are typically quenched and exhibit redder colors \citep[e.g.,][]{Wetzel2014,Rhee2017}. Instead, the blue colors and substantial stellar mass content of the W tail support a scenario in which it represents a recently accreted substructure, either still in the pre-infall phase or following an off-axis flyby of the Virgo cluster, prior to experiencing the full impact of cluster-specific environmental processes. In addition, at relative velocities of order $\sim1000$ km s$^{-1}$, typical values for galaxies interacting with the Virgo cluster, ram-pressure stripping is expected to remove the gas content of dwarf galaxies very efficiently, leading to a rapid quenching on short timescales of $\lesssim100$--200 Myr \citep{Boselli2008}. If the W tail galaxies had already passed through the dense central regions of the cluster, they would therefore be expected to appear predominantly quenched and gas-poor. The observed blue colors of these galaxies instead indicate that they have likely not yet experienced such a passage through the cluster core.

Taken together, these properties indicate that the W group is a relatively low-density system in which the stellar mass budget is dominated by a small number of massive galaxies ($M_{\ast} \gtrsim 10^{10}$~M$_{\odot}$), embedded within a much larger population of low-mass dwarf satellites. The most massive member is comparable to a Milky Way–like galaxy, while massive galaxies account for only $\sim5\%$ of the group members. The predominantly blue colors of the group galaxies suggest that internal environmental processing has been weak, with little evidence for widespread quenching prior to the interaction with the Virgo cluster. In this respect, the W group resembles nearby, dynamically young systems such as the LG or the M81 group, in which satellite galaxies can retain their gas and continue forming stars over extended periods. This picture is consistent with the idea that the W group entered the Virgo environment largely unprocessed, while the slightly bluer colors of the W tail are primarily linked to the lower average stellar mass of its members rather than to a direct effect of the interaction with the Virgo cluster.

\section{Discussion}

The results presented in the previous sections reveal that the W cloud and its associated W plume constitute one of the most dynamically and structurally complex regions in the outskirts of the Virgo cluster. Within them, we identified two spatially and dynamically linked structures, the W group and the W tail, with the former being an infalling group on its first interaction with the Virgo cluster and the latter the result of the stripping of galaxies due to this early interaction. This environment therefore provides a unique opportunity to investigate the physical mechanisms driving galaxy transformation during the hierarchical assembly of clusters.

\subsection{The interaction between the W group and the Virgo cluster}

The kinematic and spatial properties of the W tail strongly suggest that it originates from a gravitational interaction between the W group and the Virgo cluster. The coherent velocity gradient observed along the plume, together with the increase in velocity dispersion toward the cluster, indicates that the system is dynamically disturbed. This behaviour is consistent with the formation of a tidal tail or bridge-like structure, which is commonly produced when a galaxy group experiences the tidal forces of a massive cluster during infall \citep[][] {Mihos2005,Rudick2009}. The overall morphology and alignment of the W tail further support a scenario in which galaxies have been stripped from the W group as it interacts with the Virgo cluster potential.

The stellar mass budget of the system provides additional quantitative support for this interpretation. The W group has a dynamical mass of $M_{\rm dyn}\sim10^{13}$\,M$_{\odot}$ and a total stellar mass of $M_{\ast}\sim 10^{11}$\,M$_{\odot}$, consistent with a dark-matter-dominated galaxy group. The stellar mass contained in the W tail, $M_{\ast,{\rm tail}}\sim10^{10}$\,M$_{\odot}$, corresponds to $\sim10$\% of the group stellar mass, indicating that tidal stripping has been partial but already significant. Such a fraction is in line with expectations for groups undergoing their first strong tidal interaction with a cluster, rather than systems that have already experienced multiple pericentric passages \citep[e.g.,][]{Gnedin2003,Bahe2013,Vijayaraghavan2015,Rhee2017}.

The stellar population properties of the W tail galaxies further reinforce this picture. Their somewhat lower stellar masses and slightly bluer median $(g-r)$ colors compared to the group population can naturally arise from their lower typical stellar masses, since lower-mass galaxies are expected to have higher specific star formation rates and therefore bluer colors. This interpretation does not require strong tidal perturbations acting directly on individual galaxies. As shown in Appendix~\ref{app:tidal}, a simple estimate of the ratio between the tidal acceleration induced by the Virgo cluster potential and the internal gravitational acceleration of a typical dwarf galaxy at $R \gtrsim 1.5$~Mpc yields $a_{\rm tidal}/a_{\rm self} \lesssim 10^{-2}$, indicating that the direct tidal effect on individual galaxies is very small. Therefore, the interaction with the Virgo cluster is more likely to affect the W group as a whole, producing the observed large-scale tail structure, rather than significantly perturbing the internal structure or stellar populations of individual galaxies.

The predominantly blue colors of W tail galaxies suggest that these objects have not yet undergone significant environmental quenching. This argues against a backsplash origin for the W tail, as backsplash galaxies that have already passed through the cluster core are typically gas-poor and quenched \citep[e.g.,][]{Wetzel2014,Rhee2017}. Instead, the observed properties are consistent with an early-stage interaction in which tidal forces acting on the group scale dominate over hydrodynamical processes. This interpretation is further supported by the location of the W group and W tail in the outskirts of the Virgo cluster ($R > R_{200}$), where the density of the intracluster medium is relatively low and ram-pressure stripping is expected to be inefficient \citep[][]{Gunn1972,Boselli2008,Yoon2017}. At these large cluster-centric distances the direct gravitational perturbations induced by the cluster potential on individual galaxies are expected to be weak, since such effects become efficient mainly in the inner regions of massive clusters. However, the global potential of the cluster can still influence the dynamics of infalling galaxy groups, perturbing their dark-matter halos and producing large-scale tidal features. In such environments, gravitational interactions between infalling groups and the cluster potential are predicted to primarily affect the group halo and its large-scale structure rather than individual galaxies \citep[][]{Haines2015,Jaffe2015}. Together, these results identify the W tail as a tidal structure formed during an early group--cluster interaction, providing a direct observational snapshot of tidal disruption in the cluster outskirts.

We compute the gravitational interaction of the Virgo cluster on the W group halo by using the formalism of Henriksen \& Byrd (1996) presented in Appendix~\ref{app:tidal}. In this case, the ratio between the tidal acceleration exerted by the Virgo cluster and the internal gravitational acceleration of the W group adopting $M_{\rm grp}\sim10^{13}\,M_\odot$, and assuming a characteristic group radius in the range $r_{\rm grp}\sim0.2$--$0.8$ Mpc, at a distance of $\sim1.5$ Mpc yields $a_{\rm tidal}/a_{\rm self}\sim0.2$--15. For large values of the group radius ($r_{\rm grp}\sim0.8$ Mpc) the tidal perturbation becomes significantly larger than the internal gravitational acceleration of the group halo, whereas for smaller radii ($r_{\rm grp}\sim0.2$ Mpc) the central regions of the group remain largely bound. Contrary to the case of individual galaxies, the tidal perturbation of the cluster can therefore exceed the internal gravity of the outer regions of the group. This indicates that the Virgo potential can significantly perturb the W group halo at these distances.

\subsection{Dynamical state and fate of the W group system}

The spatial and kinematic properties of the W group indicate that it is a dynamically coherent structure currently located in the outskirts of the Virgo cluster. Its mean velocity of $\langle V_{LG}\rangle \simeq 2200~\mathrm{km~s^{-1}}$ and its position at a distance of $\sim17.7$~Mpc from the Milky Way place it  behind the main body of the Virgo cluster \citep{Mei2007, Cantiello2024}. The presence of a velocity gradient in projection and in three-dimensional space, together with an increase in velocity dispersion along the direction toward Virgo, suggests that the W group is being tidally perturbed by the cluster potential. This interpretation is reinforced by the existence of the W tail, a spatially elongated structure extending from the group toward the cluster, which likely traces material stripped from the group during its interaction with Virgo.

The W group–W tail system therefore captures a transitional phase where group pre–processing merges with the onset of cluster environmental influence. A simple dynamical timescale estimate is consistent with this interpretation. The W group lies at a distance of $\sim2.5$~Mpc from the Virgo cluster center and is located behind Virgo along the line of sight, and has a velocity offset of $\sim900$~km~s$^{-1}$ relative to the main cluster component. This implies a characteristic interaction timescale of order 2--3~Gyr, with plausible values in the range 1--3~Gyr when considering projection effects and velocity uncertainties. Thus, the onset of tidal perturbation --- and the formation of the W tail --- likely occurred within the past $\sim$Gyr, consistent with the W group being in an early stage of its first infall toward the Virgo cluster, during which gravitational effects act on the group halo before the system encounters the dense intracluster medium of Virgo. The fact that the W tail is observed today as a coherent spatial and kinematic structure is itself consistent with this timescale. Numerical simulations show that tidal tails and streams produced during group--cluster interactions remain identifiable for of order $\sim$1--2~Gyr before being dispersed by phase mixing and differential orbital precession \citep[e.g.,][]{Mihos2005,Rudick2009,Vijayaraghavan2015}. The survival of a well-defined W tail therefore supports a recent stripping event, in agreement with the estimated interaction timescale and a first-infall scenario.

However, the phase-space location of the W group galaxies indicates that, at present, the group as a whole is not gravitationally bound to the Virgo cluster. Most W group members lie outside the Virgo caustics, implying that the system has not yet been fully captured by the cluster potential. This raises the question of the ultimate fate of the W group: whether it will eventually be accreted by Virgo or whether it has already experienced a close passage and will not fall back onto the cluster.

One plausible scenario is that the W group is on a first infall orbit and will become gravitationally bound to Virgo at a later stage. In this case, the observed tidal perturbations and the formation of the W tail would represent the early stages of group disruption prior to pericentric passage. Numerical simulations show that groups approaching massive clusters can experience significant tidal stripping even before becoming formally bound, particularly during high-velocity encounters \citep[e.g.,][]{Vijayaraghavan2015,Rhee2017}. The coherent velocity structure and alignment of the W tail are consistent with this picture. Additional support for an early-stage interaction comes from the stellar mass distribution of galaxies in the system. The four most massive galaxies of the W group ($M_{\ast} > 10^{10}$~M$_{\odot}$) remain bound to the group core, while the W tail is predominantly populated by lower-mass dwarf galaxies. This suggests that the present interaction has preferentially affected the least massive systems, which are more weakly bound to the group potential. The fact that the most massive galaxies have not been removed from the group halo indicates that the gravitational perturbation induced by the Virgo cluster has not yet been strong enough to disrupt the deepest part of the group potential. This behaviour is naturally expected during a first encounter between a galaxy group and a cluster, before the system experiences a deeper pericentric passage.

An alternative scenario is that the W group has already undergone an off-axis flyby interaction with the Virgo cluster and is now moving away from the cluster, remaining unbound. In this case, the W tail would trace material stripped during this close passage, with the earliest stripped galaxies already bound to Virgo and the more recently stripped material still following the group trajectory. The clear segregation observed in velocity and phase space between bound and unbound W tail galaxies supports the existence of such an interaction sequence.

Regardless of its ultimate fate, the W group demonstrates that significant tidal processing can occur beyond the virial radius, leaving long-lived imprints on both group and cluster populations.

\subsection{Environmental mechanisms in the outskirts of the cluster}

Galaxy evolution in the outskirts of clusters is governed by a combination of gravitational and hydrodynamical processes, whose relative importance depends on local environment and orbital history \citep[e.g.,][]{Boselli2006,Moore1996}. In the case of the W group and its associated W tail, the available evidence indicates that gravitational mechanisms currently dominate the interaction with the Virgo cluster.

The coherent spatial and kinematic structure of the W tail strongly supports a tidal origin, consistent with the progressive stripping of low-mass galaxies from the W group as it interacts with the cluster potential. The smooth velocity gradients observed both in projection and in three-dimensional space, together with the elongated morphology of the plume pointing toward Virgo, are characteristic signatures of tidally disturbed galaxy groups \citep[e.g.,][]{Mihos2005,Vijayaraghavan2015,Rhee2017}. Such features can arise well outside the cluster virial radius, where tidal forces from the global potential already become significant \citep[e.g.,][]{Ghigna1998,Behroozi2014}.

The stellar population properties further constrain the nature of the environmental mechanisms at work. Galaxies in the W tail are systematically less massive than those in the W group and exhibit slightly bluer median $(g-r)$ colors. However, both populations lie within the blue cloud, indicating that they have not yet experienced significant quenching. The slightly bluer colors of the W tail galaxies can be naturally explained by their lower stellar masses, since low-mass galaxies are expected to have higher specific star formation rates and therefore bluer colors. This suggests that the observed color differences between the W group and W tail do not necessarily require strong environmental perturbations acting on individual galaxies. The absence of a significant quenched population also indicates that hydrodynamical processes associated with the intracluster medium remain inefficient at the current location of the system \citep[e.g.,][]{Gunn1972,Yoon2017}.

The W group itself is dominated by a small number of massive galaxies ($M_\ast \gtrsim 10^{10}$~M$_\odot$), with the bulk of the population consisting of low-mass dwarfs, resulting in a high dwarf fraction ($DF \gtrsim 0.9$). This demographic closely resembles the galaxy groups identified in the Virgo infall region by \cite{ChoqueChallapa2026}, which are likewise characterized by high dwarf fractions and minimal signs of strong pre-processing. Together, these properties support a scenario in which the W group entered the Virgo environment as a relatively unevolved, low-density system.

\subsection{The buildup of the dwarf galaxy population in the Virgo cluster}

Dwarf galaxies ($M_{B} > -18$ or $M_{\ast} < 10^{9}$~M$_{\odot}$) constitute the most numerous galaxy population in the local Universe, particularly in galaxy clusters, where the faint-end slope of the luminosity function is steeper than $\alpha \simeq -1$ \citep[e.g.,][]{Popesso2005,Agulli2014}. The relative abundance of dwarfs compared to massive galaxies varies with environment and redshift \citep[e.g.,][]{DeLucia2007,Bildfell2012,Barkhouse2009,Rude2020}, indicating that cluster dwarf populations are assembled through multiple evolutionary pathways.

The origin of cluster dwarf galaxies remains debated. They may form in situ, be transformed by cluster-specific mechanisms, or be accreted as part of galaxy groups with varying degrees of pre-processing \citep[e.g.,][]{Moore1996,Mastropietro2005,Boselli2008,Fujita2004,DeLucia2012}. Recent studies suggest that a substantial fraction of dwarfs formed outside clusters and were accreted at late times \citep[e.g.,][]{RomeroGomez2023,RomeroGomez2024a,RomeroGomez2024b,ChoqueChallapa2026}.

In this context, the W group and its associated W tail provide evidence for an additional accretion channel. The W group is a low-density system with little internal pre-processing, while the W tail consists of low-mass, star-forming dwarfs that have not yet been quenched.

The W group--W tail system therefore illustrates how poor galaxy groups can contribute actively to the buildup of the cluster dwarf population, adding a population of initially unquenched dwarf galaxies whose subsequent evolution will likely be influenced by environmental processes as they progressively enter the cluster environment.

\section{Conclusions}

We presented a detailed analysis of the southern region of the Virgo cluster, focusing on the W cloud and its associated W plume in the cluster outskirts. By combining spatial, kinematic, and stellar population information, we have characterized the nature of this system and its role in the ongoing assembly of the Virgo cluster. Our main conclusions are as follows:

\begin{enumerate}

\item The W cloud is not a filament seen head-on, as previously suggested. Instead, it is a collection of galaxies dominated by a compact system, the W group, which is currently interacting with the Virgo cluster.

\item A tail of galaxies, the W tail, traces a path between the W group and the Virgo cluster. This structure is most plausibly interpreted as the result of tidal stripping able to remove galaxies from the gravitational potential of the W group during its interaction with the Virgo cluster. This tail constitutes a dynamically coherent structure whose velocity distribution reveals two distinct components: a low-velocity population of galaxies already bound to the Virgo potential, and a high-velocity population that is still transitioning between the group and the cluster. The continuous variation of velocity and three-dimensional distance along the tail provides a rare and direct observational snapshot of cluster growth in progress.

\item The spatial configuration and kinematic properties of the W tail strongly indicate a tidal origin. The observed velocity gradients and the increase in velocity dispersion toward the cluster are consistent with the formation of a tidal tail generated by the interaction between the W group and the Virgo cluster. Numerical simulations predict that such tidal debris can remain coherent for $\sim$1--2~Gyr before being dispersed by phase mixing and orbital precession. The existence of a well-defined W tail is therefore fully consistent with a recent group--cluster interaction. These results demonstrate that significant tidal stripping of group galaxies can occur well beyond the cluster virial radius.

\item The stellar mass and color properties of the W tail galaxies show that the stripped population is dominated by low-mass dwarfs that remain star-forming and lie within the blue cloud. The slightly bluer colors observed in the W tail compared to the W group can be naturally explained by the somewhat lower stellar masses of the galaxies in the tail, since lower-mass galaxies are expected to have higher specific star formation rates and therefore bluer colors. This interpretation does not require invoking enhanced star formation triggered by tidal interactions. Instead, the stellar populations of the W tail galaxies are consistent with systems that entered the cluster environment largely unprocessed, retaining their gas and ongoing star formation prior to experiencing the stronger environmental effects that typically operate deeper inside the cluster.

\end{enumerate}

The spatial, kinematic, and evolutionary continuity between the W group, the W tail, and the Virgo cluster provides compelling evidence that the hierarchical growth of clusters proceeds not only through the accretion of individual galaxies, but also through the infall and partial disruption of galaxy groups \citep{Tully1984}. In this process, tidal interactions play a key role in redistributing low-mass galaxies and shaping the emerging cluster population.

Overall, the W group--W tail system offers a unique and well-resolved laboratory for studying the early stages of environmental transformation and the buildup of the dwarf galaxy population in nearby clusters. By capturing an active group--cluster interaction in real time, this system provides strong observational constraints on how clusters assemble their low-mass galaxy populations through multiple, complementary accretion pathways.

\begin{acknowledgements}

  JALA acknowledge financial support provided by the Spanish Ministerio de Ciencia, Innovación y Universidades (MICIU) through the project PID2023-153342NB-I00 / 10.13039/501100011033. SZ acknowledges the financial support provided by the Governments of Spain and Arag\'on through their general budgets and the Fondo de Inversiones de Teruel, the Aragonese Government through the Research Group E16\_23R, and the Spanish Ministry of Science and Innovation and the European Union - NextGenerationEU through the Recovery and Resilience Facility project ICTS-MRR-2021-03-CEFCA. VC acknowledges the support provided by ANID through the FONDECYT grant no. 11250723. LM and JALA acknowledges a support grant from the Joint Committee ESO-Government of Chile (ORP 058/2023). Funding for the Sloan Digital Sky Survey IV has been provided by the Alfred P. Sloan Foundation, the U.S. Department of Energy Office of Science, and the Participating Institutions. SDSS acknowledges support and resources from the Center for High-Performance Computing at the University of Utah. The SDSS web site is www.sdss4.org. SDSS is managed by the Astrophysical Research Consortium for the Participating Institutions of the SDSS Collaboration including the Brazilian Participation Group, the Carnegie Institution for Science, Carnegie Mellon University, Center for Astrophysics Harvard \& Smithsonian (CfA), the Chilean Participation Group, the French Participation Group, Instituto de Astrofísica de Canarias, The Johns Hopkins University, Kavli Institute for the Physics and Mathematics of the Universe (IPMU) / University of Tokyo, the Korean Participation Group, Lawrence Berkeley National Laboratory, Leibniz Institut für Astrophysik Potsdam (AIP), Max-Planck-Institut für Astronomie (MPIA Heidelberg), Max-Planck-Institut für Astrophysik (MPA Garching), Max-Planck-Institut für Extraterrestrische Physik (MPE), National Astronomical Observatories of China, New Mexico State University, New York University, University of Notre Dame, Observatório Nacional / MCTI, The Ohio State University, Pennsylvania State University, Shanghai Astronomical Observatory, United Kingdom Participation Group, Universidad Nacional Autónoma de México, University of Arizona, University of Colorado Boulder, University of Oxford, University of Portsmouth, University of Utah, University of Virginia, University of Washington, University of Wisconsin, Vanderbilt University, and Yale University. 
\end{acknowledgements}

\bibliography{bibliografia}

\begin{appendix}
\section{Tidal effect of the Virgo cluster on individual dwarf galaxies and at group scales}
\label{app:tidal}

In this Appendix we estimate the strength of the tidal perturbation exerted by the gravitational potential of the Virgo cluster on individual galaxies belonging to the W group and W tail. This calculation quantifies whether the cluster potential can significantly perturb individual galaxies at the present cluster-centric distances due to off-axis flyby encounters similar to those described in the present work.

The tidal acceleration induced by an external body of mass $M$ at a distance $R$ on a galaxy of characteristic size $r_{\rm gal}$ can be approximated as (e.g., Henriksen \& Byrd 1996; Cortese et al. 2007)

\begin{equation}
a_{\rm tidal} \simeq \frac{2GM_{\rm cl} r_{\rm gal}}{R^3},
\end{equation}

where $M_{\rm cl}$ is the mass of the cluster and $R$ is the distance from the cluster center.

The internal gravitational acceleration binding the galaxy can be approximated by

\begin{equation}
a_{\rm self} \simeq \frac{G m_{\rm gal}}{r_{\rm gal}^2},
\end{equation}

where $m_{\rm gal}$ is the mass of the galaxy.

The relative importance of the tidal perturbation can therefore be estimated from the ratio

\begin{equation}
\frac{a_{\rm tidal}}{a_{\rm self}} \simeq
2 \frac{M_{\rm cl}}{m_{\rm gal}}
\left(\frac{r_{\rm gal}}{R}\right)^3 .
\end{equation}

We adopt representative values appropriate for the Virgo cluster and for dwarf galaxies in the W group and tail structures. For the Virgo cluster we assume a total mass $M_{\rm cl} \sim 5\times10^{14}\,M_{\odot}$ \citep[see][]{McLaughlin1999} and a characteristic cluster-centric distance $R \gtrsim 1.5\,{\rm Mpc}$ ( $\sim R_{200}$ of the Virgo cluster) for the W group and W tail galaxies. For a typical dwarf galaxy we adopt a stellar mass $m_{\rm gal} \sim 10^{9}\,M_{\odot}$ and a characteristic size $r_{\rm gal} \sim 3\,{\rm kpc}$.

Using these values we obtain

\begin{equation}
\left(\frac{r_{\rm gal}}{R}\right)^3 =
\left(\frac{3\,{\rm kpc}}{1500\,{\rm kpc}}\right)^3
\simeq 8\times10^{-9}.
\end{equation}

Combining the above expressions yields

\begin{equation}
\frac{a_{\rm tidal}}{a_{\rm self}}
\simeq 2 \times
\frac{5\times10^{14}}{10^{9}}
\times 8\times10^{-9}
\sim 8\times10^{-3}.
\end{equation}

Therefore, the tidal acceleration induced by the Virgo cluster potential represents less than $\sim1\%$ of the internal gravitational acceleration binding a typical dwarf galaxy.

This simple estimate indicates that, at cluster-centric distances of order $R \gtrsim 1.5$ Mpc, the direct tidal effect of the Virgo cluster on individual galaxies is very small. Consequently, the cluster potential alone is unlikely to significantly perturb the internal structure of these galaxies. Therefore, the observed properties of the galaxies in the W group and W tail are unlikely to be primarily driven by the direct tidal interaction with the gravitational potential of the Virgo cluster.

We can perform a similar estimate to evaluate the tidal influence of the Virgo cluster on an entire galaxy group located at similar cluster-centric distances. In this case the relevant comparison is between the tidal acceleration exerted by the cluster across the group and the internal gravitational acceleration binding the group itself.

The tidal acceleration across a group of characteristic size $r_{\rm grp}$ at a distance $R$ from the cluster center can be written as

\begin{equation}
a_{\rm tidal} \simeq \frac{2GM_{\rm cl} r_{\rm grp}}{R^3},
\end{equation}

while the internal gravitational acceleration of the group is approximately

\begin{equation}
a_{\rm grp} \simeq \frac{G M_{\rm grp}}{r_{\rm grp}^2},
\end{equation}

where $M_{\rm grp}$ is the total mass of the galaxy group.

The ratio between the external tidal field and the self-gravity of the group is therefore

\begin{equation}
\frac{a_{\rm tidal}}{a_{\rm grp}}
\simeq
2 \frac{M_{\rm cl}}{M_{\rm grp}}
\left(\frac{r_{\rm grp}}{R}\right)^3 .
\end{equation}

We adopt a representative group mass of $M_{\rm grp} \sim 10^{13}\,M_{\odot}$ similar to our W group, and the same Virgo cluster mass and distance used above ($M_{\rm cl} = 5\times10^{14}\,M_{\odot}$ and $R = 1.5\,{\rm Mpc}$). We explore three characteristic group sizes for the W group: $r_{\rm grp}=0.2$, $0.5$, and $0.8$ Mpc.

For $r_{\rm grp}=0.2$ Mpc:

\begin{equation}
\left(\frac{r_{\rm grp}}{R}\right)^3 =
\left(\frac{0.2}{1.5}\right)^3
\simeq 2.4\times10^{-3}
\end{equation}

\begin{equation}
\frac{a_{\rm tidal}}{a_{\rm grp}}
\simeq
2 \times 50 \times 2.4\times10^{-3}
\approx 0.24 .
\end{equation}

For $r_{\rm grp}=0.5$ Mpc:

\begin{equation}
\left(\frac{r_{\rm grp}}{R}\right)^3 =
\left(\frac{0.5}{1.5}\right)^3
\simeq 3.7\times10^{-2}
\end{equation}

\begin{equation}
\frac{a_{\rm tidal}}{a_{\rm grp}}
\simeq
2 \times 50 \times 3.7\times10^{-2}
\approx 3.7 .
\end{equation}

For $r_{\rm grp}=0.8$ Mpc:

\begin{equation}
\left(\frac{r_{\rm grp}}{R}\right)^3 =
\left(\frac{0.8}{1.5}\right)^3
\simeq 1.5\times10^{-1}
\end{equation}

\begin{equation}
\frac{a_{\rm tidal}}{a_{\rm grp}}
\simeq
2 \times 50 \times 1.5\times10^{-1}
\approx 15 .
\end{equation}

These simple estimates indicate that while the Virgo cluster tidal field is negligible for the internal structure of individual dwarf galaxies, its effect on galaxy groups can be substantial. For compact groups ($r_{\rm grp}\sim0.2$ Mpc) the tidal field represents a moderate perturbation, while for more extended groups ($r_{\rm grp}\gtrsim0.5$ Mpc) the cluster tidal field can become comparable to or even exceed the group's self-gravity.

This suggests that the Virgo cluster potential may significantly affect the dynamical evolution of infalling galaxy groups, as our W group, potentially contributing to their tidal distortion, partial disruption, or the generation of tidal features during their interaction with the cluster environment.

\end{appendix}

\end{document}